\documentclass[namedreferences]{solarphysics}

\usepackage[optionalrh,solaromanenum]{spr-sola-addons} 
\usepackage{graphicx}                    
\usepackage{amssymb}                    
\usepackage{color}                       

\begin{document}

\begin{article}

\begin{opening}

	\title{GLE and Sub-GLE Redefinition in the Light of High-Altitude Polar Neutron Monitors}

%
\author[addressref={af1,af2},corref,email={stepan.poluianov@oulu.fi}]
{\inits{S.V.~}\fnm{S.V.~}\lnm{Poluianov}\orcid{orcid.org/0000-0002-9119-4298}}
\author[addressref={af1,af2},corref,email={}]
{\inits{}\fnm{I.G.~}\lnm{Usoskin}\orcid{orcid.org/0000-0001-8227-9081}}
\author[addressref={af1,af2},corref,email={}]
{\inits{}\fnm{A.L.~}\lnm{Mishev}\orcid{orcid.org/0000-0002-7184-9664}}
\author[addressref={af3},corref,email={}]
{\inits{}\fnm{M.A.~}\lnm{Shea}}
\author[addressref={af3},corref,email={}]
{\inits{}\fnm{D.F.~}\lnm{Smart}}

%
	\runningauthor{S. Poluianov \textit{et al.}}
\runningtitle{GLE and Sub-GLE Redefinition...}

\address[id=af1]{Space Climate Research Unit, University of Oulu, Finland}
\address[id=af2]{Sodankyl\"a Geophysical Observatory, University of Oulu, Finland}
\address[id=af3]{SSSRC, 100 Tennyson Avenue, Nashua, NH 03062, USA}

\begin{abstract}
The conventional definition of ground level enhancement (GLE)\break events 
requires detection of solar energetic particles (SEP) by at least two 
differently located neutron monitors.
Some places are exceptionally suitable for ground-based detection of 
SEP -- high-elevation polar regions with 
negligible geomagnetic and reduced atmospheric energy/rigidity cutoffs.
At present, there are two neutron-monitor stations in such locations on 
the Antarctic plateau:
SOPO/SOPB (at Amundsen--Scott station, 2835 m elevation) and 
DOMC/\break DOMB (at Concordia station, 3233 m elevation).
Since 2015, when the DOMC/\break DOMB station started continuous operation, 
a relatively weak SEP event, not detected by sea-level neutron-monitor 
stations, was registered by both SOPO/\break SOPB and DOMC/DOMB and accordingly 
classified as a GLE.
This would lead to a distortion of the homogeneity of the historic
GLE list and the corresponding statistics.
To address this issue, we propose to modify the GLE definition so that 
it keeps the homogeneity:
A GLE event is registered when there are near-time coincident and
statistically significant enhancements of the count rates of at least two
differently located neutron monitors 
including at least one neutron monitor near sea level and 
a corresponding enhancement in the proton flux measured by
a space-borne instrument(s).
Relatively weak SEP events registered only by 
high-altitude polar neutron monitors, but with no response from
cosmic-ray stations at sea level, can be classified as sub-GLEs.
\end{abstract}

%

\end{opening}

%
\section{Introduction}
\label{s:intro}

A nearly simultaneous enhancement of the count rates of several ground-based 
cosmic-ray detectors, neutron monitors (NM), or ionization chambers
\citep[e.g.][]{Forbush_PhR1946, Simpson_ICRC1990},
caused by solar energetic particles (SEP) is known as
a ground-level enhancement (GLE) event.
Observations of GLEs provide key information about the high-energy portion
(above several hundred MeV nuc$^{-1}$) of strong SEP events, 
which cannot be continuously monitored by space-borne 
instruments \citep{Tylka_ICRC2009}.
We note that the modern space-borne particle spectrometers such as
the alpha magnetic spectrometer AMS-02 \citep{AMS02_PRL2015} and 
payload for antimatter matter exploration and light-nuclei astrophysics
\citep[PAMELA,][]{PAMELA_ApPh2007} are 
not well suited to study SEP events
because in low orbits they spend most of the time in regions
with high geomagnetic cutoff.

The GLE dataset spans high-energy SEP events over almost seven solar cycles 
providing sufficient basis for statistical studies 
\citep[e.g.][]{Gopalswamy_EPS2014, Raukunen_JSWSC2017, Vainio_AA2017}.
Since the beginning of systematic ground-based
measurements of cosmic rays, over 70 GLEs have been ``officially'' registered 
so far (see the International GLE Database 
\href{http://gle.oulu.fi}{\texttt{gle.oulu.fi}}).
However, in 2015 there was a SEP event 
that was clearly observed by the South Pole (SOPO/SOPB) and 
Dome C (DOMC/DOMB)
stations, but it was hardly distinguishable in the count rates of
other NMs.
Based on the current GLE definition, it would have been identified as a GLE,
but on the other hand, it would not, if the DOMC/DOMB station had not been operational.
Accordingly, this can compromise the homogeneity of the GLE dataset and 
introduce a bias in the statistics.
Here we propose a revision of the GLE definition to keep the GLE dataset 
homogeneous.

\section{The Present GLE Definition and its Limitation}

The definition of a GLE was proposed by the cosmic-ray community in the 1970s
and is commonly understood as the following:
%
\begin{quote}
\textit{A GLE event is registered when there are near-time coincident and
statistically significant enhancements of the count rates of
at least two differently located NMs.
}
\end{quote}
We note that previously there was only one high-latitude 
high-altitude NM station, the South Pole, 
and a SEP event must have been recorded at the sea-level to be 
identified as a GLE.
However, with the installation of DOMC/DOMB, the situation has changed significantly.
The sensitivity of a NM to cosmic rays depends on the rigidity/energy threshold 
of detectable particles,
defined by the geomagnetic- and atmospheric-shielding effects.
The geomagnetic shielding is quantified in terms of 
the geomagnetic cutoff rigidity, which varies from almost
zero in the polar region to 15\,--\,17 GV at the geomagnetic Equator 
\citep[e.g.][]{Smart2009, Nevalainen2013}.
In addition, there is the atmospheric cutoff, implying that 
a primary particle must possess some minimal energy to be able to 
initiate the atmospheric cascade and be registered on the ground.
It depends on the thickness of the atmosphere above the location,
being greatest \citep[$\approx 430$ MeV nuc$^{-1}$, see][]{Grieder2001,Dorman2004} at 
sea level and decreasing with altitude.
The atmospheric cutoff dominates shielding in the polar regions, and 
elevation of a cosmic-ray detector above sea level 
reduces the atmospheric attenuation of cosmic-ray cascades significantly.
Accordingly, cosmic-ray detectors located at high altitude in 
the polar region possess higher sensitivity to
low-energy cosmic rays 
relative to ones at near-sea-level and/or non-polar locations.

\begin{figure} 
\centerline{\includegraphics[width=0.8\textwidth]{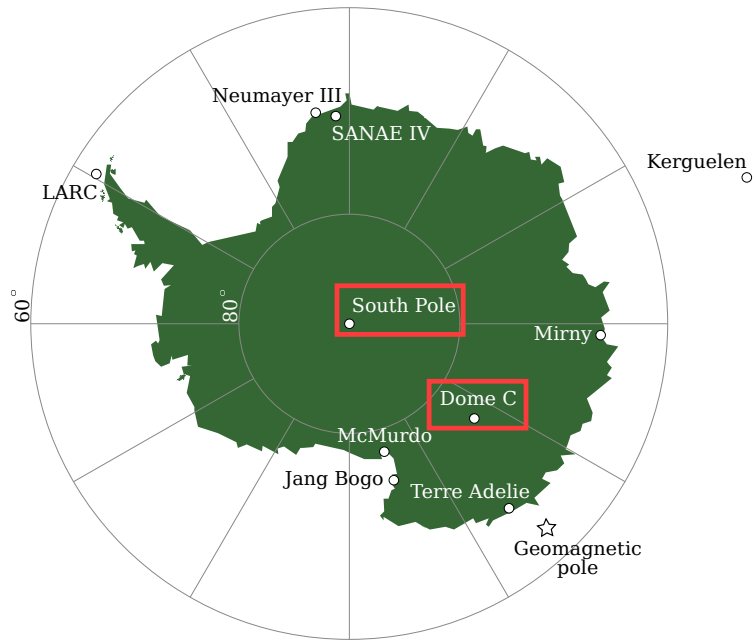}}
	\caption{Antarctic neutron-monitor stations operational in 2017. 
	Red boxes indicate the high-altitude stations.
	The map is adopted from \copyright Gringer/Wikimedia Commons/GFDL.}
\label{f:map}
\end{figure}

There are two regions with these properties in the world: 
the top of the Greenland ice sheet (3205 m above the sea level (asl))
and the Antarctic plateau (average elevation of about 3000 m asl).
Although there is no high-altitude cosmic-ray station in Greenland, 
two NMs are located on the Antarctic plateau:
at South Pole and at Dome C (Figure \ref{f:map}).

The Amundsen--Scott US research station at the geographical South Pole
(90$^\circ$S, elevation of 2835 m asl, the nominal barometric pressure 690 hPa) 
had a neutron monitor in operation since 1964 
\citep[\texttt{http://nmdb.eu}]{Evenson_ICRC2011}.
The setup consists of two types of cosmic-ray detectors:
a standard NM-64 instrument with a lead layer for multiplication of neutrons,
and a so-called ``bare'' or lead-free NM without the lead layer 
\citep[e.g.][]{Vashenyuk2007}.
The South Pole standard and ``bare'' NMs are denoted as SOPO and SOPB, 
respectively.

A more recent neutron monitor, called ``Dome C'', is located at 
the Franco--Italian research station Concordia
(75$^\circ06'$S, 123$^\circ23'$E, 3233 m asl, 650 hPa).
It is also equipped with the two types of neutron monitors denoted as 
DOMC (standard) and DOMB (``bare'').
The Dome C cosmic-ray station started operation in early 2015 
\citep{Poluianov_JASS2015}.

\begin{figure} 
\centerline{\includegraphics[width=1\textwidth,clip=]{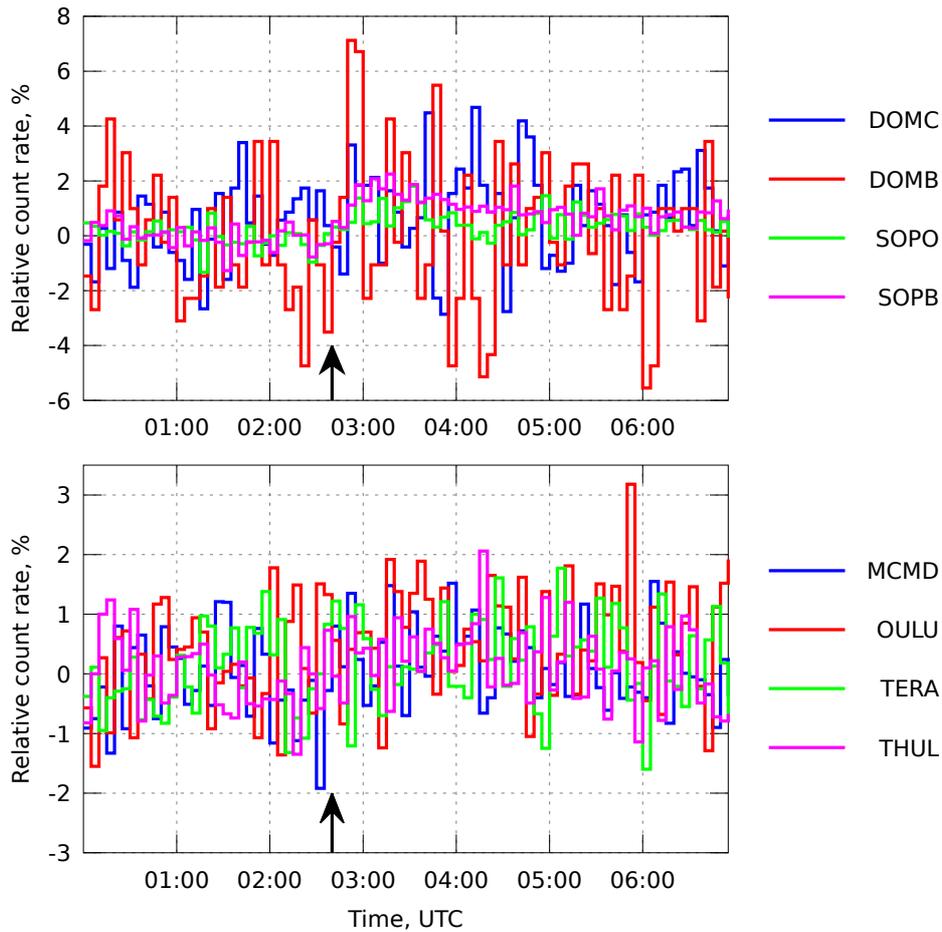}}
	\caption{Solar-energetic-particle event on 29 October 2015 in 
	neutron-monitor data (see the color legends).
	Dome C (DOMC, DOMB) and South Pole (SOPO, SOPB) NMs are shown in 
	the upper panel, while McMurdo (MCMD), Oulu (OULU),
	Terre Adelie (TERA), and Thule (THUL) NMs plotted in the lower one.
	Arrows indicate the onset of the event.
	Data are from the International GLE Database
	(\href{http://gle.oulu.fi}{\texttt{gle.oulu.fi}}).}
\label{f:subgle}
\end{figure}

Before 2015 the global NM network had only one high-altitude 
polar station (SOPO/SOPB).
All GLEs were registered by instruments including at least one 
located near the sea level, implying
that SEP events causing GLEs had sufficiently large flux of particles with 
energy above the full atmospheric cutoff of $\approx 1$ GV 
($\approx 430$ MeV nuc$^{-1}$).
The installation of the DOMC/DOMB detector in 2015 has lead to a situation when 
a SEP event with much lower flux of energetic particles can be formally 
classified as a GLE.
There have already been several SEP events that have been registered by 
only those instruments and confirmed by data from spacecrafts,
but these events have not been
detected by other NMs at near sea-level elevations.
One example is an event of 29 October 2015 that took place around 
02:40 UT, as shown in Figure \ref{f:subgle}.
A distinguishable signal was recorded only by SOPO/SOPB and DOMC/DOMB
(left panel), while other polar sea-level NMs did not register any 
significant response.
Thus, if applying the commonly used GLE definition, this event 
might have been
considered as a GLE, while it is obvious that it would not have been 
counted earlier, without DOMC/DOMB data.
The SEP event on 29 October 2015 is different from other GLE events because of 
its lower intensity and much softer spectrum compared with a weak 
``official'' GLE \citep{Mishev_JSWSC2017}.
The ratio of SEP peak fluxes, computed based on five-minute NM data 
during the maximum phase of the event using spectrum reconstructions 
by \citet{Mishev2014, Mishev_SPh2016}, for this event
relative to the weak GLE on 17 May 2012 is 0.25, 0.13, and 0.05 for 
the energy ranges $>200$, $>300$, and $>500$ MeV, respectively.
One can see that the event of 29 October 2015 was not only weaker, but also 
much softer than the GLE on 17 May 2012.
It was also under the alert threshold of the GLE alert system 
\citep{Souvatzoglou_ASR2009}.
This introduces a bias to the definition of GLE, since the NM network is 
more sensitive to GLE now than ever before, and therefore, 
some of the events observed nowadays would not have been recorded 
if they had occurred in the past. 
Accordingly, we believe that applying the ``classical'' GLE definition 
now can significantly distort the homogeneity of the existing list.

When a small increase is measured in a high-altitude 
polar NM, it is normal to look for a corresponding increase in 
sea-level NMs; however, when investigating what might initially 
appear to be a small increase in the counting rate of 
a sea level NM, care must be taken to differentiate between
the normal daily variation, random fluctuation, and 
increase from solar particles.
In addition, we acknowledge the uncertainty in exact timing of 
GLE detection caused by different magnetic connection of NM stations 
during the initial highly anisotropic phase of a particle event.

\section{Proposed Definitions of GLE and Sub-GLE}

To address the issue described above, we suggest modifying 
the GLE definition,
namely, to fix the condition that the event should be detected near 
sea level, implying that the corresponding 
SEP event has sufficient flux of particles with the energy 
exceeding the full atmospheric cutoff.

We propose to formulate the revised GLE definition as follows:
\begin{quote}
\textit{
A GLE event is registered when there are near-time coincident and
statistically significant enhancements of the count rates of at least two
differently located neutron monitors 
including at least one neutron monitor near sea level and 
a corresponding enhancement in the proton flux measured by
a space-borne instrument(s).}
\end{quote}
By the term ``near the sea level'', we mean altitude not higher
than approximately a quarter of the mean attenuation
length of the nucleonic component of 
the cosmic-ray cascade \citep[e.g.][]{Grieder2001, Dorman2004} 
above the sea level, viz. $\approx 30$ g cm$^{-2}$.
This corresponds to $\approx 1000$ g cm$^{-2}$ in total atmospheric depth
or $\approx 300$ m asl in altitude.
NMs located at higher altitudes possess a notably 
different response function due to the cascade attenuation effects 
\citep{Clem_SSR2000, Fluekiger_ICRC2007}.
It is important to note that the proposed definition does not affect 
the existing list of ``classic'' GLEs,
because all of them have been detected by at least one sea-level NM
\citep[or ionization chambers for the four events before 1950,][]{Forbush_PhR1950}.

However, it would be incorrect to ignore completely SEP events 
registered by only high-altitude polar NMs, without a significant
response from the near sea-level NMs. 
Such events (as the one of 29 October 2015 discussed above) can be called sub-GLEs.
Although this term is not very precise, it is already used in 
the literature \citep[e.g.][]{Atwell_ICES2015, Vainio_AA2017} to 
classify events with a reduced, opposed to a ``classic'' GLE, proton energy range.
For sub-GLE, we propose the following definition:
\begin{quote}
\textit{
	A sub-GLE event is registered when there are near-time coincident and
statistically significant enhancements of the count rates of at least two
differently located high-elevation neutron monitors
and a corresponding enhancement in the proton flux measured by a space-borne
instrument(s), but no statistically significant enhancement in
the count rates of neutron monitors near sea level.
}
\end{quote}
We note that this definition does not contradict 
the one proposed by \citet{Atwell_ICES2015}.

\section{Summary}

The installation of the second high-altitude polar cosmic-ray station 
DOMC/ DOMB,
in addition to the long-existing South Pole SOPO/SOPB station, 
has improved the sensitivity of the global NM network, 
so now it can detect SEP events below the full atmospheric cutoff, 
which would not been recorded previously and classified as GLEs.
This distorts the homogeneity of the list of GLE
and may affect studies based on the GLE long-term occurrence rate.
In order to address this issue, we have suggested modifying
the GLE definition by requiring that at least one of the NMs recording 
the event should be located near sea level.
The new definition does not affect the list of already registered events,
but rejects some weaker SEP events of Cycle 24 as being GLEs.
SEP events detected by only high-altitude polar NMs, 
with no significant response of NMs near sea level, can be 
called sub-GLEs.
The corresponding formal definition of such events is proposed.

%
\begin{acks}
The work was supported by the projects of the Academy of Finland
Centre of Excellence ReSoLVE (No. 272157), CRIPA and CRIPA-X (No. 304435),
and Finnish Antarctic Research Program (FINNARP).
Operation of the DOMC/DOMB NMs was possible due to support of
the French-Italian Concordia Station. 
French financial support and field logistic supplies for 
the year-round campaigns at Dome C were provided by 
the Institut Polaire Fran\c{c}ais-Paul Emile Victor (IPEV) through program n903.
We acknowledge Askar Ibragimov for the support of 
the International GLE database (\href{http://gle.oulu.fi}{\texttt{gle.oulu.fi}})
and are thankful to the worldwide neutron-monitor database 
(\href{http://www.nmdb.eu}{\texttt{nmdb.eu}}), which is a product of an EU Project.
We thank Marc Duldig, Erwin Fl\"uckiger, John Humble, and 
Roger Pyle for valuable discussions.
\end{acks}

\section*{Disclosure of Potential Conflicts of Interest}
The authors declare that they have no conflicts of interest.


\begin{thebibliography}{23}
\ifx\bisbn     \undefined \def\bisbn  #1{ISBN #1}\fi
\ifx\binits    \undefined \def\binits#1{#1}\fi
\ifx\bauthor   \undefined \def\bauthor#1{#1}\fi
\ifx\batitle   \undefined \def\batitle#1{#1}\fi
\ifx\bjtitle   \undefined \def\bjtitle#1{\textit{#1}}\fi
\ifx\bvolume   \undefined \def\bvolume#1{\textbf{#1}}\fi
\ifx\byear     \undefined \def\byear#1{#1}\fi
\ifx\bissue    \undefined \def\bissue#1{#1}\fi
\ifx\bfpage    \undefined \def\bfpage#1{#1}\fi
\ifx\blpage    \undefined \def\blpage #1{#1}\fi
\ifx\burl      \undefined \def\burl#1{\textsf{#1}}\fi
\ifx\href      \undefined \def\href#1#2{\textsf{#2}}\fi
\ifx\betal     \undefined \def\betal{\textit{et al.}}\fi
\ifx\bctitle   \undefined \def\bctitle#1{#1}\fi
\ifx\beditor   \undefined \def\beditor#1{#1}\fi
\ifx\bbtitle   \undefined \def\bbtitle#1{\textit{#1}}\fi
\ifx\bedition  \undefined \def\bedition#1{#1}\fi
\ifx\bseriesno \undefined \def\bseriesno#1{\textbf{#1}}\fi
\ifx\blocation \undefined \def\blocation#1{#1}\fi
\ifx\bsertitle \undefined \def\bsertitle#1{\textit{#1}}\fi
\ifx\bsnm      \undefined \def\bsnm#1{#1}\fi
\ifx\bsuffix   \undefined \def\bsuffix#1{#1}\fi
\ifx\bparticle \undefined \def\bparticle#1{#1}\fi
\ifx\barticle  \undefined \def\barticle#1{}\fi
\ifx\binstitute  \undefined \def\binstitute#1{#1}\fi
\ifx\bpublisher  \undefined \def\bpublisher#1{#1}\fi
\ifx\doiurl    \undefined
  \def\doiurl#1{\href{http://dx.doi.org/#1}{\textsf{DOI}}}\fi
\ifx\arxivurl  \undefined
  \def\arxivurl#1{\href{http://arxiv.org/abs/#1}{\textsf{arXiv}}}\fi
\ifx\adsurl    \undefined
  \def\adsurl#1{\href{http://adsabs.harvard.edu/abs/#1}{\textsf{ADS}}}\fi
\ifx\botherref \undefined \def\botherref#1{}\fi
\ifx\url       \undefined \def\url#1{\textsf{#1}}\fi
\ifx\bchapter  \undefined \def\bchapter#1{}\fi
\ifx\bbook     \undefined \def\bbook#1{}\fi
\ifx\bcomment  \undefined \def\bcomment#1{#1}\fi
\ifx\oauthor   \undefined \def\oauthor#1{#1}\fi
\ifx\citeauthoryear \undefined\def \citeauthoryear#1{#1}\fi
\ifx\endbibitem\undefined \def\endbibitem{}\fi
\ifx\bconflocation  \undefined \def\bconflocation#1{#1} \fi

\bibitem[\protect\citeauthoryear{Aguilar \textit{et~al.}}{2015}]{AMS02_PRL2015}
\begin{barticle}
\bauthor{\bsnm{Aguilar}, \binits{M.}},
\bauthor{\bsnm{Aisa}, \binits{D.}},
\bauthor{\bsnm{Alpat}, \binits{B.}},
\bauthor{\bsnm{Alvino}, \binits{A.}},
\bauthor{\bsnm{Ambrosi}, \binits{G.}},
\bauthor{\bsnm{Andeen}, \binits{K.}}, \betal:
\byear{2015},
\batitle{{Precision Measurement of the Proton Flux in Primary Cosmic Rays from
  Rigidity 1 GV to 1.8 TV with the Alpha Magnetic Spectrometer on the
  International Space Station}}.
\bjtitle{Phys. Rev. Lett.}
\bvolume{114},
\bfpage{171103}.
\doiurl{10.1103/PhysRevLett.114.171103}.
\end{barticle}
\endbibitem

\bibitem[\protect\citeauthoryear{{Atwell}
  \textit{et~al.}}{2015}]{Atwell_ICES2015}
\begin{bchapter}
\bauthor{\bsnm{{Atwell}}, \binits{W.}},
\bauthor{\bsnm{{Tylka}}, \binits{A.J.}},
\bauthor{\bsnm{{Dietrich}}, \binits{W.}},
\bauthor{\bsnm{{Rojdev}}, \binits{K.}},
\bauthor{\bsnm{{Matzkind}}, \binits{C.}}:
\byear{2015},
\bctitle{{Sub-GLE Solar Particle Events and the Implications for
  Lightly-Shielded Systems Flown During an Era of Low Solar Activity}}.
In: \bbtitle{Int. Conf. Environ. Sys.},
\bsertitle{Lunar Planet. Sci. Conf. Proc.}
\burl{ntrs.nasa.gov/search.jsp?R=20150009484}.
\end{bchapter}
\endbibitem

\bibitem[\protect\citeauthoryear{Clem and Dorman}{2000}]{Clem_SSR2000}
\begin{barticle}
\bauthor{\bsnm{Clem}, \binits{J.M.}},
\bauthor{\bsnm{Dorman}, \binits{L.I.}}:
\byear{2000},
\batitle{Neutron monitor response functions}.
\bjtitle{Space Sci. Rev.}
\bvolume{93},
\bfpage{335}.
\doiurl{10.1023/A:1026508915269}.
\adsurl{2000SSRv...93..335C}.
\end{barticle}
\endbibitem

\bibitem[\protect\citeauthoryear{Dorman}{2004}]{Dorman2004}
\begin{bbook}
\bauthor{\bsnm{Dorman}, \binits{L.}}:
\byear{2004},
\bbtitle{Cosmic rays in the earth's atmosphere and underground},
\bpublisher{Kluwer},
\blocation{Dordrecht}.
\bisbn{1-4020-2071-6}.
\end{bbook}
\endbibitem

\bibitem[\protect\citeauthoryear{{Evenson}
  \textit{et~al.}}{2011}]{Evenson_ICRC2011}
\begin{bchapter}
\bauthor{\bsnm{{Evenson}}, \binits{P.}},
\bauthor{\bsnm{{Bieber}}, \binits{J.}},
\bauthor{\bsnm{{Clem}}, \binits{J.}},
\bauthor{\bsnm{{Pyle}}, \binits{R.}}:
\byear{2011},
\bctitle{{South Pole Neutron Monitor Lives Again}}.
In: \bbtitle{Proc. Int. Cosmic Ray Conf.}
\bseriesno{11},
\bfpage{459}.
\doiurl{10.7529/ICRC2011/V11/0622}.
\adsurl{2011ICRC...11..459E}.
\end{bchapter}
\endbibitem

\bibitem[\protect\citeauthoryear{{Fl{\"u}ckiger}
  \textit{et~al.}}{2008}]{Fluekiger_ICRC2007}
\begin{bchapter}
\bauthor{\bsnm{{Fl{\"u}ckiger}}, \binits{E.O.}},
\bauthor{\bsnm{{Moser}}, \binits{M.R.}},
\bauthor{\bsnm{{Pirard}}, \binits{B.}},
\bauthor{\bsnm{{B{\"u}tikofer}}, \binits{R.}},
\bauthor{\bsnm{{Desorgher}}, \binits{L.}}:
\byear{2008},
\bctitle{{A parameterized neutron monitor yield function for space weather
  applications}}.
In: \beditor{\bsnm{Caballero}, \binits{R.}},
\beditor{\bsnm{D'Olivo}, \binits{J.C.}},
\beditor{\bsnm{Medina-Tanco}, \binits{G.}},
\beditor{\bsnm{Nellen}, \binits{L.}},
\beditor{\bsnm{Sánchez}, \binits{F.A.}},
\beditor{\bsnm{Valdés-Galicia}, \binits{J.F.}} (eds.)
\bbtitle{Proc. 30th Int. Cosmic Ray Conf.}
\bseriesno{1},
\bfpage{289}.
\adsurl{2008ICRC....1..289F}.
\end{bchapter}
\endbibitem

\bibitem[\protect\citeauthoryear{{Forbush}}{1946}]{Forbush_PhR1946}
\begin{barticle}
\bauthor{\bsnm{{Forbush}}, \binits{S.E.}}:
\byear{1946},
\batitle{{Three Unusual Cosmic-Ray Increases Possibly Due to Charged Particles
  from the Sun}}.
\bjtitle{Phys. Rev.}
\bvolume{70},
\bfpage{771}.
\doiurl{10.1103/PhysRev.70.771}.
\adsurl{1946PhRv...70..771F}.
\end{barticle}
\endbibitem

\bibitem[\protect\citeauthoryear{{Forbush}, {Stinchcomb}, and
  {Schein}}{1950}]{Forbush_PhR1950}
\begin{barticle}
\bauthor{\bsnm{{Forbush}}, \binits{S.E.}},
\bauthor{\bsnm{{Stinchcomb}}, \binits{T.B.}},
\bauthor{\bsnm{{Schein}}, \binits{M.}}:
\byear{1950},
\batitle{{The Extraordinary Increase of Cosmic-Ray Intensity on November 19,
  1949}}.
\bjtitle{Phys. Rev.}
\bvolume{79},
\bfpage{501}.
\doiurl{10.1103/PhysRev.79.501}.
\adsurl{1950PhRv...79..501F}.
\end{barticle}
\endbibitem

\bibitem[\protect\citeauthoryear{{Gopalswamy}
  \textit{et~al.}}{2014}]{Gopalswamy_EPS2014}
\begin{barticle}
\bauthor{\bsnm{{Gopalswamy}}, \binits{N.}},
\bauthor{\bsnm{{Xie}}, \binits{H.}},
\bauthor{\bsnm{{Akiyama}}, \binits{S.}},
\bauthor{\bsnm{{M{\"a}kel{\"a}}}, \binits{P.A.}},
\bauthor{\bsnm{{Yashiro}}, \binits{S.}}:
\byear{2014},
\batitle{{Major solar eruptions and high-energy particle events during solar
  cycle 24}}.
\bjtitle{Earth Planet. Space}
\bvolume{66},
\bfpage{104}.
\doiurl{10.1186/1880-5981-66-104}.
\adsurl{2014EP\%26S...66..104G}.
\end{barticle}
\endbibitem

\bibitem[\protect\citeauthoryear{{Grieder}}{2001}]{Grieder2001}
\begin{bbook}
\bauthor{\bsnm{{Grieder}}, \binits{P.K.F.}}:
\byear{2001},
\bbtitle{{Cosmic Rays at Earth}},
\bpublisher{Elsevier Science},
\blocation{Amsterdam}.
\adsurl{2001cre..book.....G}.
\end{bbook}
\endbibitem

\bibitem[\protect\citeauthoryear{{Mishev} and {Usoskin}}{2016}]{Mishev_SPh2016}
\begin{barticle}
\bauthor{\bsnm{{Mishev}}, \binits{A.}},
\bauthor{\bsnm{{Usoskin}}, \binits{I.}}:
\byear{2016},
\batitle{{Analysis of the Ground-Level Enhancements on 14 July 2000 and 13
  December 2006 Using Neutron Monitor Data}}.
\bjtitle{Solar Phys.}
\bvolume{291},
\bfpage{1225}.
\doiurl{10.1007/s11207-016-0877-2}.
\adsurl{2016SoPh..291.1225M}.
\end{barticle}
\endbibitem

\bibitem[\protect\citeauthoryear{Mishev, Kocharov, and
  Usoskin}{2014}]{Mishev2014}
\begin{barticle}
\bauthor{\bsnm{Mishev}, \binits{A.L.}},
\bauthor{\bsnm{Kocharov}, \binits{L.G.}},
\bauthor{\bsnm{Usoskin}, \binits{I.G.}}:
\byear{2014},
\batitle{Analysis of the ground level enhancement on 17 may 2012 using data
  from the global neutron monitor network}.
\bjtitle{J. Geophys. Res.: Space Phys.}
\bvolume{119}(\bissue{2}),
\bfpage{670}.
\doiurl{10.1002/2013JA019253}.
\end{barticle}
\endbibitem

\bibitem[\protect\citeauthoryear{{Mishev, Alexander}, {Poluianov, Stepan}, and
  {Usoskin, Ilya}}{2017}]{Mishev_JSWSC2017}
\begin{barticle}
\bauthor{\bsnm{{Mishev, Alexander}}},
\bauthor{\bsnm{{Poluianov, Stepan}}},
\bauthor{\bsnm{{Usoskin, Ilya}}}:
\byear{2017},
\batitle{Assessment of spectral and angular characteristics of sub-gle events
  using the global neutron monitor network}.
\bjtitle{J. Space Weather Space Clim.}
\bvolume{7},
\bfpage{A28}.
\doiurl{10.1051/swsc/2017026}.
\end{barticle}
\endbibitem

\bibitem[\protect\citeauthoryear{Nevalainen, Usoskin, and
  Mishev}{2013}]{Nevalainen2013}
\begin{barticle}
\bauthor{\bsnm{Nevalainen}, \binits{J.}},
\bauthor{\bsnm{Usoskin}, \binits{I.}},
\bauthor{\bsnm{Mishev}, \binits{A.}}:
\byear{2013},
\batitle{Eccentric dipole approximation of the geomagnetic field: Application
  to cosmic ray computations}.
\bjtitle{Adv. Space Res.}
\bvolume{52}(\bissue{1}),
\bfpage{22}.
\doiurl{10.1016/j.asr.2013.02.020}.
\end{barticle}
\endbibitem

\bibitem[\protect\citeauthoryear{Picozza
  \textit{et~al.}}{2007}]{PAMELA_ApPh2007}
\begin{barticle}
\bauthor{\bsnm{Picozza}, \binits{P.}},
\bauthor{\bsnm{Galper}, \binits{A.M.}},
\bauthor{\bsnm{Castellini}, \binits{G.}},
\bauthor{\bsnm{Adriani}, \binits{O.}},
\bauthor{\bsnm{Altamura}, \binits{G.}},
\bauthor{\bsnm{Ambriola}, \binits{M.}}, \betal:
\byear{2007},
\batitle{{PAMELA --- A payload for antimatter matter exploration and
  light-nuclei astrophysics}}.
\bjtitle{Astropart. Phys.}
\bvolume{27}(\bissue{4}),
\bfpage{296 }.
\doiurl{https://doi.org/10.1016/j.astropartphys.2006.12.002}.
\end{barticle}
\endbibitem

\bibitem[\protect\citeauthoryear{{Poluianov}
  \textit{et~al.}}{2015}]{Poluianov_JASS2015}
\begin{barticle}
\bauthor{\bsnm{{Poluianov}}, \binits{S.}},
\bauthor{\bsnm{{Usoskin}}, \binits{I.}},
\bauthor{\bsnm{{Mishev}}, \binits{A.}},
\bauthor{\bsnm{{Moraal}}, \binits{H.}},
\bauthor{\bsnm{{Kruger}}, \binits{H.}},
\bauthor{\bsnm{{Casasanta}}, \binits{G.}},
\bauthor{\bsnm{{Traversi}}, \binits{R.}},
\bauthor{\bsnm{{Udisti}}, \binits{R.}}:
\byear{2015},
\batitle{{Mini Neutron Monitors at Concordia Research Station, Central
  Antarctica}}.
\bjtitle{J. Astron. Space Sci.}
\bvolume{32},
\bfpage{281}.
\doiurl{10.5140/JASS.2015.32.4.281}.
\adsurl{2015JASS...32..281P}.
\end{barticle}
\endbibitem

\bibitem[\protect\citeauthoryear{Raukunen
  \textit{et~al.}}{2017}]{Raukunen_JSWSC2017}
\begin{botherref}
\oauthor{\bsnm{Raukunen}, \binits{O.}},
\oauthor{\bsnm{Vainio}, \binits{R.}},
\oauthor{\bsnm{Tylka}, \binits{A.J.}},
\oauthor{\bsnm{Dietrich}, \binits{W.F.}},
\oauthor{\bsnm{Jiggens}, \binits{P.}},
\oauthor{\bsnm{Heynderickx}, \binits{D.}},
\oauthor{\bsnm{Dierckxsens}, \binits{M.}},
\oauthor{\bsnm{Crosby}, \binits{N.}},
\oauthor{\bsnm{Ganse}, \binits{U.}},
\oauthor{\bsnm{Siipola}, \binits{R.}}:
2017,
{Two solar proton fluence models based on ground level enhancement
  observations}.
\textit{J. Space Weather Space Clim.}
submitted.
\end{botherref}
\endbibitem

\bibitem[\protect\citeauthoryear{{Simpson}}{1990}]{Simpson_ICRC1990}
\begin{bchapter}
\bauthor{\bsnm{{Simpson}}, \binits{J.A.}}:
\byear{1990},
\bctitle{{Astrophysical Phenomena Discovered by Cosmic Ray and Solar Flare
  Ground Level Events: The Early Years}}.
In: \bbtitle{Proc. Int. Cosmic Ray Conf.}
\bseriesno{12},
\bfpage{187}.
\adsurl{1990ICRC...12..187S}.
\end{bchapter}
\endbibitem

\bibitem[\protect\citeauthoryear{Smart and Shea}{2009}]{Smart2009}
\begin{barticle}
\bauthor{\bsnm{Smart}, \binits{D.F.}},
\bauthor{\bsnm{Shea}, \binits{M.A.}}:
\byear{2009},
\batitle{Fifty years of progress in geomagnetic cutoff rigidity
  determinations}.
\bjtitle{Adv. Space Res.}
\bvolume{44}(\bissue{10}),
\bfpage{1107}.
\doiurl{10.1016/j.asr.2009.07.005}.
\end{barticle}
\endbibitem

\bibitem[\protect\citeauthoryear{Souvatzoglou
  \textit{et~al.}}{2009}]{Souvatzoglou_ASR2009}
\begin{barticle}
\bauthor{\bsnm{Souvatzoglou}, \binits{G.}},
\bauthor{\bsnm{Mavromichalaki}, \binits{H.}},
\bauthor{\bsnm{Sarlanis}, \binits{C.}},
\bauthor{\bsnm{Mariatos}, \binits{G.}},
\bauthor{\bsnm{Belov}, \binits{A.}},
\bauthor{\bsnm{Eroshenko}, \binits{E.}},
\bauthor{\bsnm{Yanke}, \binits{V.}}:
\byear{2009},
\batitle{{Real-time GLE alert in the ANMODAP Center for December 13, 2006}}.
\bjtitle{Adv. Space Res.}
\bvolume{43}(\bissue{4}),
\bfpage{728 }.
\bcomment{Solar Extreme Events: Fundamental Science and Applied Aspects}.
\doiurl{10.1016/j.asr.2008.09.018}.
\end{barticle}
\endbibitem

\bibitem[\protect\citeauthoryear{Tylka and Dietrich}{2009}]{Tylka_ICRC2009}
\begin{bchapter}
\bauthor{\bsnm{Tylka}, \binits{A.J.}},
\bauthor{\bsnm{Dietrich}, \binits{W.F.}}:
\byear{2009},
\bctitle{A new and comprehensive analysis of proton spectra in ground-level
  enhanced (gle) solar particle events}.
In: \bbtitle{Proc. 31th Int. Cosmic Ray Conf.},
\bconflocation{Lodz}.
\end{bchapter}
\endbibitem

\bibitem[\protect\citeauthoryear{{Vainio}
  \textit{et~al.}}{2017}]{Vainio_AA2017}
\begin{barticle}
\bauthor{\bsnm{{Vainio}}, \binits{R.}},
\bauthor{\bsnm{{Raukunen}}, \binits{O.}},
\bauthor{\bsnm{{Tylka}}, \binits{A.J.}},
\bauthor{\bsnm{{Dietrich}}, \binits{W.F.}},
\bauthor{\bsnm{{Afanasiev}}, \binits{A.}}:
\byear{2017},
\batitle{{Why is solar cycle 24 an inefficient producer of high-energy particle
  events?}}
\bjtitle{Astron. Astrophys.}
\bvolume{604},
\bfpage{A47}.
\doiurl{10.1051/0004-6361/201730547}.
\adsurl{2017A\%26A...604A..47V}.
\end{barticle}
\endbibitem

\bibitem[\protect\citeauthoryear{Vashenyuk, Balabin, and
  Stoker}{2007}]{Vashenyuk2007}
\begin{barticle}
\bauthor{\bsnm{Vashenyuk}, \binits{E.V.}},
\bauthor{\bsnm{Balabin}, \binits{Y.V.}},
\bauthor{\bsnm{Stoker}, \binits{P.H.}}:
\byear{2007},
\batitle{Responses to solar cosmic rays of neutron monitors of a various
  design}.
\bjtitle{Adv. Space Res.}
\bvolume{40}(\bissue{3}),
\bfpage{331}.
\doiurl{10.1016/j.asr.2007.05.018}.
\end{barticle}
\endbibitem

\end{thebibliography}

\end{article} 
\end{document}